\documentclass[showpacs,twocolumn,preprintnumbers,amsmath,amssymb]{revtex4}
\usepackage{amsmath,amsfonts,latexsym,amssymb,graphicx,graphics,epsfig,subfigure,color,makeidx}
\usepackage{xcolor,diagbox}
\usepackage{multirow}
\usepackage[colorlinks,linkcolor=blue,anchorcolor=blue,citecolor=green,urlcolor=blue]{hyperref}

\usepackage[latin1]{inputenc}

\newcommand {\be}    {\begin{equation}}
\newcommand {\ee}    {\end{equation}}
\newcommand {\beq}   {\begin{eqnarray}}
\newcommand {\eeq}   {\end{eqnarray}}

\begin{document}
\title{Comment on \textquotedblleft Traversable Wormholes in General Relativity\textquotedblright}

\author{Yong-Qiang Wang\footnote{yqwang@lzu.edu.cn}}
\author{Shao-Wen Wei\footnote{weisw@lzu.edu.cn}}
\author{Yu-Xiao Liu\footnote{liuyx@lzu.edu.cn, corresponding author}}

\affiliation{$^{a}$Lanzhou Center for Theoretical Physics, Key Laboratory of Theoretical Physics of Gansu Province,
School of Physical Science and Technology, Lanzhou University, Lanzhou 730000, People's Republic of China\\
             $^{b}$Institute of Theoretical Physics $\&$ Research Center of Gravitation, Lanzhou University, Lanzhou 730000, People's Republic of China}

\begin{abstract}  
In the  letter titled  \textquotedblleft Traversable Wormholes in General Relativity\textquotedblright
 [Phys. Rev. Lett. {\bf 128},  091104  (2022)],
R. A. Konoplya and A. Zhidenko have constructed  an asymmetric wormhole solution, which is  not symmetric about the throat and  is compounded  from smooth gravitational and  charged Dirac fields. However, the authors  have  claimed that a  physically relevant condition on the throat is imposed to lead to  no gravitational force experienced by a stationary observer at the throat.
 In this comment, we point out that the above condition is unnecessary.
\end{abstract}
\maketitle



The letter titled ``Traversable Wormholes in General Relativity'' \cite{Konoplya:2021hsm}
is an interesting work that investigated the model of  a Maxwell field coupling to  two Dirac fields {in General Relativity}, and  {sought to} establish a family of asymptotically flat, asymmetric  {(about the throat)} and traversable wormhole solutions. These solutions describe {wormholes compounded} from smooth gravitational and Dirac fields,
which is based on the previous researchs in Refs.~\cite{Blazquez-Salcedo:2020czn,Blazquez-Salcedo:2021udn}.

The authors of Ref.~\cite{Konoplya:2021hsm} claimed that a physically relevant condition on the throat $x=0$  {should  be imposed}.  {This condition was chosen as}~\cite{Konoplya:2021hsm}
\begin{equation}\label{equ1}
	N'(0)=0,
\end{equation}
where the function $N(x)$ is related to the metric coefficient $g_{tt}$.  {The above condition} could lead to no gravitational force experienced by a stationary observer at the throat.

The numerical method used by the authors  is the shooting method. According to the asymptotic behavior of the  {gravitational field} equations at  the throat, ones can know that  if one only fixes the  {values} of $F(0), G(0), N(0)$, and $B'(0)$, it is not enough to determine the first-order derivatives of these functions, and there are still redundant free parameters, which will lead to failure of the shooting method. So, one needs to fix the value of  $N'(0)$. However, considering that the traversable wormhole solutions  {are generally} asymmetric at the  {throat, we should have $N'(0)\neq0$, which
means that the wormhole should be nonsymmetric  {about} the throat \cite{Bolokhov:2021fil}.
Of course, it is also possible that the condition $N'(0)=0$ {also} leads to an asymmetric solution. {But this is just a special case}.
In addition,  {it is reasonable for an asymmetric wormhole that there is a} gravitational force experienced by a stationary observer at the throat.
Based on the above two reasons, we {propose} that  it is  necessary to extend the condition in Eq.~(\ref{equ1}) to
\begin{equation}\label{equ2}
	N'(0)=\text{Constant},
\end{equation}
where the constant may be limited in a range, while each of them corresponds to a  certain wormhole solution.

Besides the shooting method,   the Newton-Raphson  method is also feasible to solve the coupled system of nonlinear ordinary differential equations.
We also  numerically solve the equations of motion (30) in Ref.~\cite{Konoplya:2021hsm}.
We first solve the {equations of motion} in $0\leq x \leq1$ with the boundary conditions at infinities:
\begin{eqnarray}\label{equ3}
	   F(1)=G(1)=0,~~~ B(1) =1,~~~\text{at}~~~ x=1,
\end{eqnarray}
and
\begin{eqnarray}\label{equ3}
  F(0)&=&f_i,~~~ B(0)=0,   ~~~N(0)=n_i, \nonumber\\
  U(0)&=&\omega, ~~~ W(0)=w_i,
~~~\text{at}~~~ x=0,
\end{eqnarray}
where $f_i$, $n_i$, $\omega$, and $w_i$ are constants.
After numerically solving the equations of motion with the above boundary conditions, we can obtain the solutions of the functions $F(x)$, $G(x)$, $N(x)$,  $B(x)$, and   $W(x)$  in $0\leq x \leq1$. Besides, the value of $N'(0)$ can also be determined. Then, in the range of $-1\leq x \leq 0$, we could continue to adopt the Newton-Raphson method or use the shooting method to solve the equations of motion, and look for smooth solutions.

\begin{table*}
\begin{tabular}{|c|c|c|c|c|c|c|c|c|c|c|}
\hline
$b_i$            &           $g_i$ & $f_i$   &      $n_i/\sigma_+$     &   $\sigma_-/\sigma_+$    &    $Q_+/r_0$    &   $Q_-/r_0$    &   $M_+/r_0$   &$M_-/r_0$   &  $\omega r_0/\sigma_+$  &      $N'(0)$
\\
\hline
$	 0.289865  $ & $ 0.0049816 $ & $  0.033000 $ & $  0.0521623$ & $   1.62879  $ & $ 0.978917  $ & $  0.977803$ & $    0.977428$ & $    0.977230$ & $  -0.120251$ & $  -0.0622616$\\
$		0.289043$ & $    0.0177181   $ & $     0.025000   $ & $0.0491175  $ & $ 1.21799  $ & $     0.976157  $ & $     0.974928  $ & $       0.974764   $ & $       0.973952   $ & $ -0.115032   $ & $  -0.0263646$	\\
$		 0.288946   $ & $ 0.0256779        $ & $   0.016095  $ & $    0.0448804  $ & $    1.00440    $ & $     0.975806  $ & $ 0.974565   $ & $  0.974637   $ & $   0.973344   $ & $    -0.105347   $ & $    0$	\\
$		 0.289748 $ & $   0.0329166$ & $   0     $ & $ 0.0345718 $ & $   0.72252$ & $  0.978405$ & $    0.977286 $ & $  0.977742 $ & $    0.975932   $ & $ -0.080254  $ & $  0.0421234$	\\
$		 0.291968    $ & $     0.0310424  $ & $  -0.028000 $ & $  0.0173830  $ & $  0.24981  $ & $  0.985702  $ & $   0.984627$ & $	0.985574    $ & $  0.982341   $ & $     -0.034964      $ & $ 0.1299027 $	\\
\hline
\end{tabular}
\caption{The families of wormhole solutions with the parameters $qr_0=0.03$ and $\mu r_0=0.2$.}\label{tabl1}
\end{table*}

In TAB.~\ref{tabl1}, we list some typical results of the values of the parameters describing the wormhole configurations.
From the last  column, one can see that each wormhole with a set of parameters  has a different value of $N'(0)$.
Except for that $N'(0)$ is   equal to zero {for} $\omega r_0/\sigma_+= -0.105347$ in the third row of the table, the rest of the wormhole solutions {has a deviation from the case of} $N'(0)=0$.  Comparing with the result of TAB. I given in Ref.~\cite{Konoplya:2021hsm}, we conjecture that  when there is a big gap between  $\sigma_-$ and $\sigma_+$, the solutions with $N'(0) \neq 0$ can exist.
Meanwhile, in FIG.~\ref{fig1}, we show a typical result of the distribution of the wormhole configuration as a function of $x$ with the parameters given in the last row of TAB. \ref{tabl1}.  In
the bottom left panel, we can find that the value of $N'(0)=0.1299027$ is not equal to zero, and the minimum value of $N(x)$ is located at the red point $x=-0.0815$.

\begin{figure}
\resizebox{\linewidth}{!}{\includegraphics{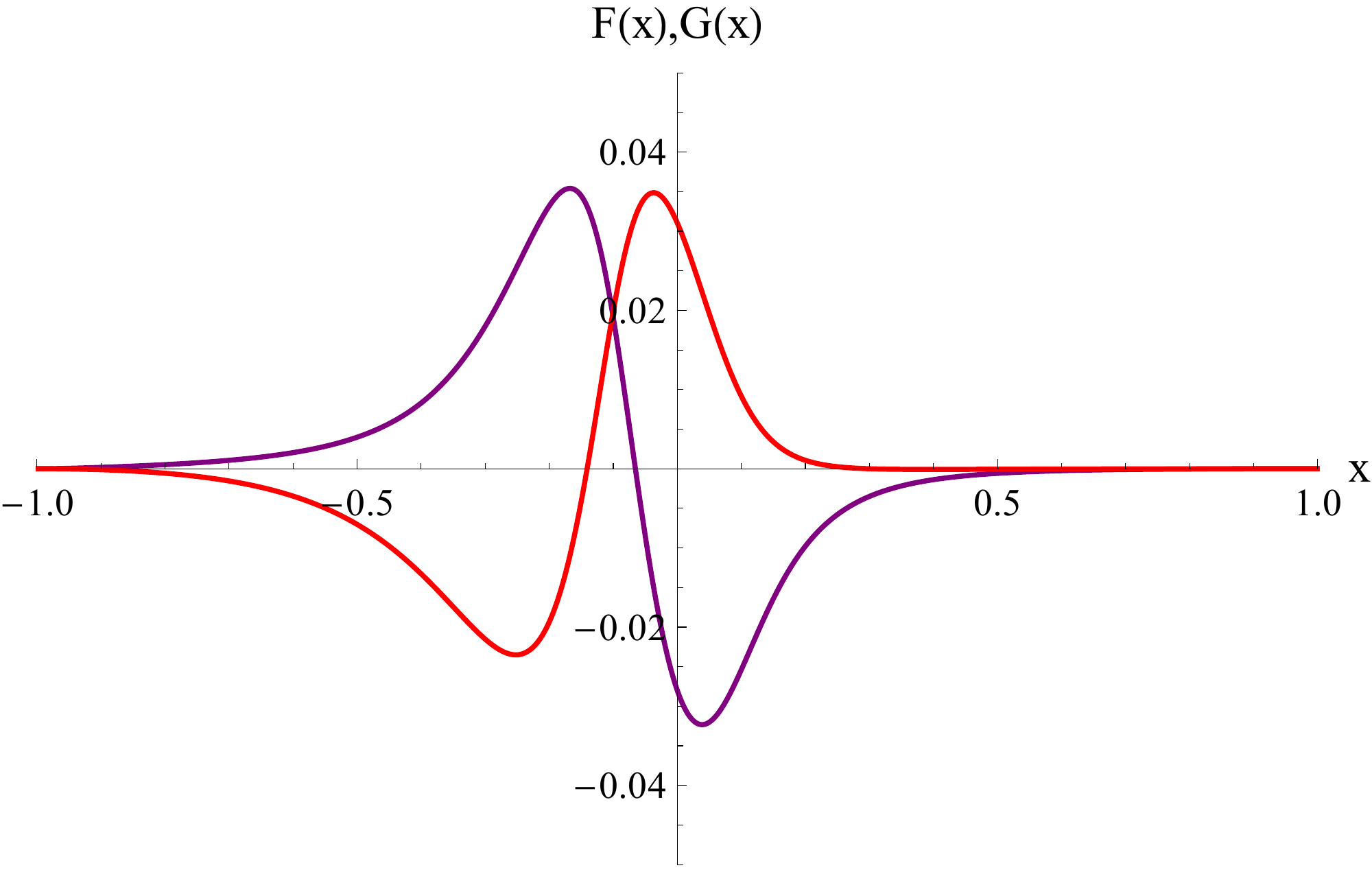}\includegraphics{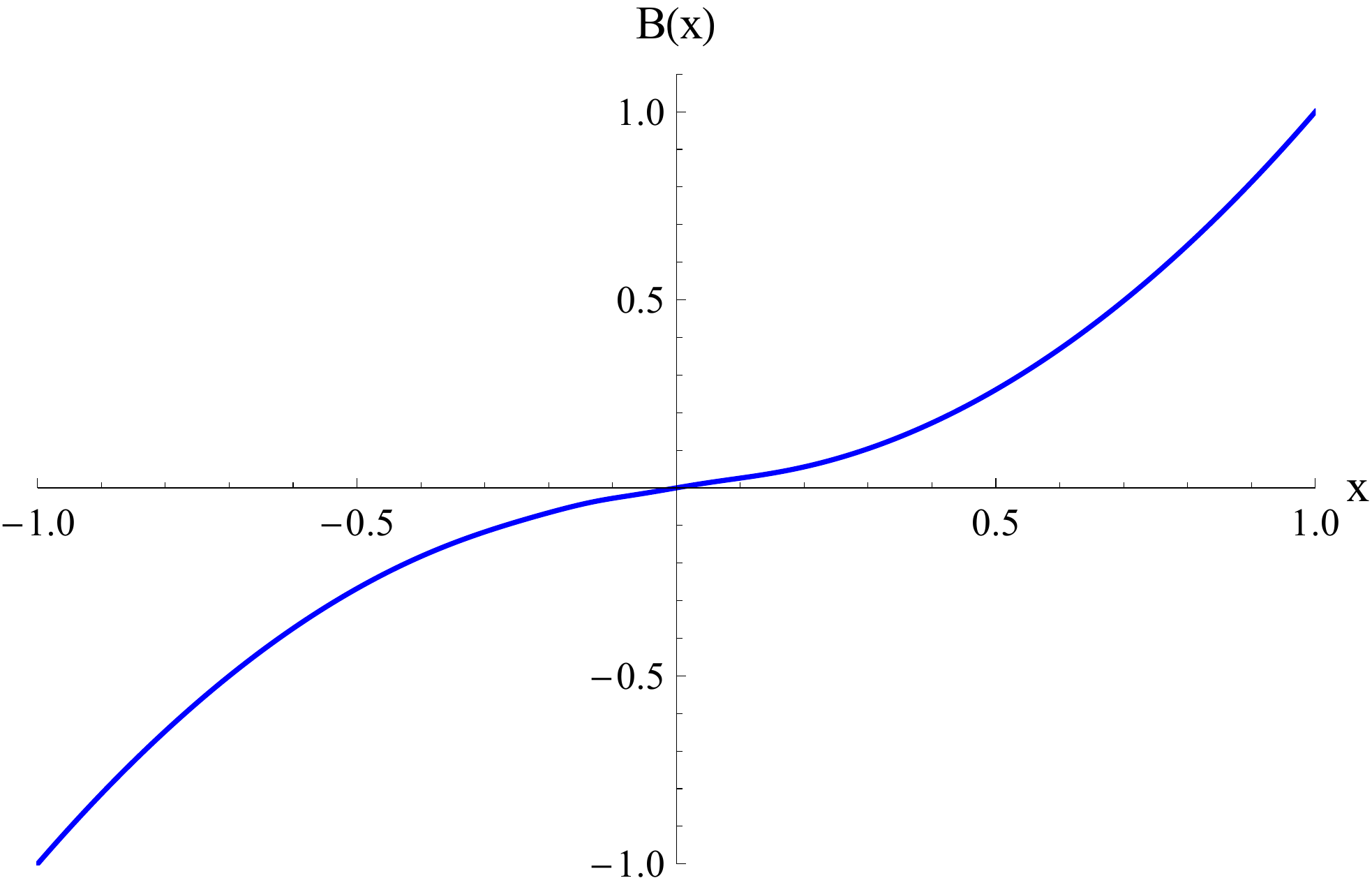}}
\resizebox{\linewidth}{!}{\includegraphics{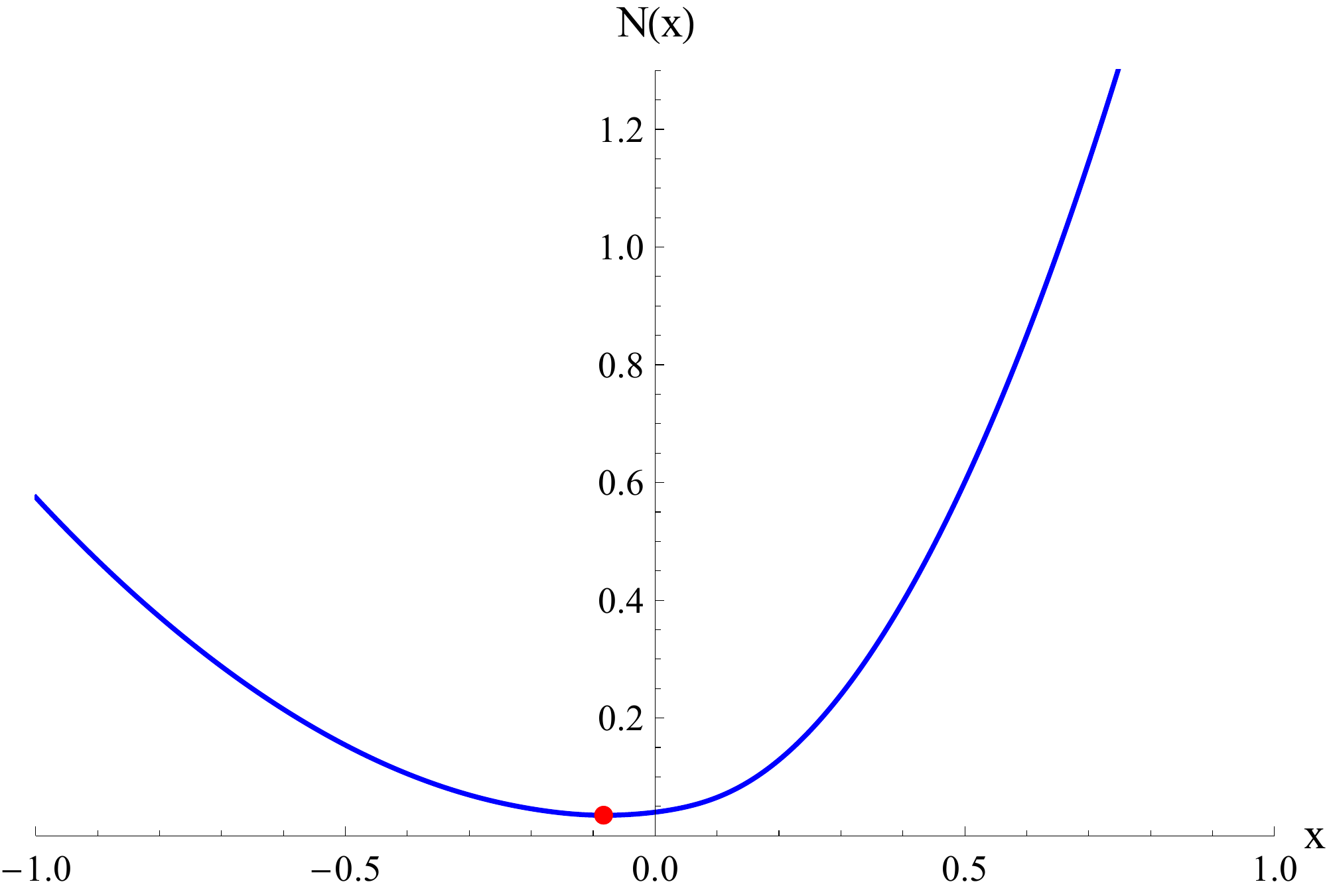}\includegraphics{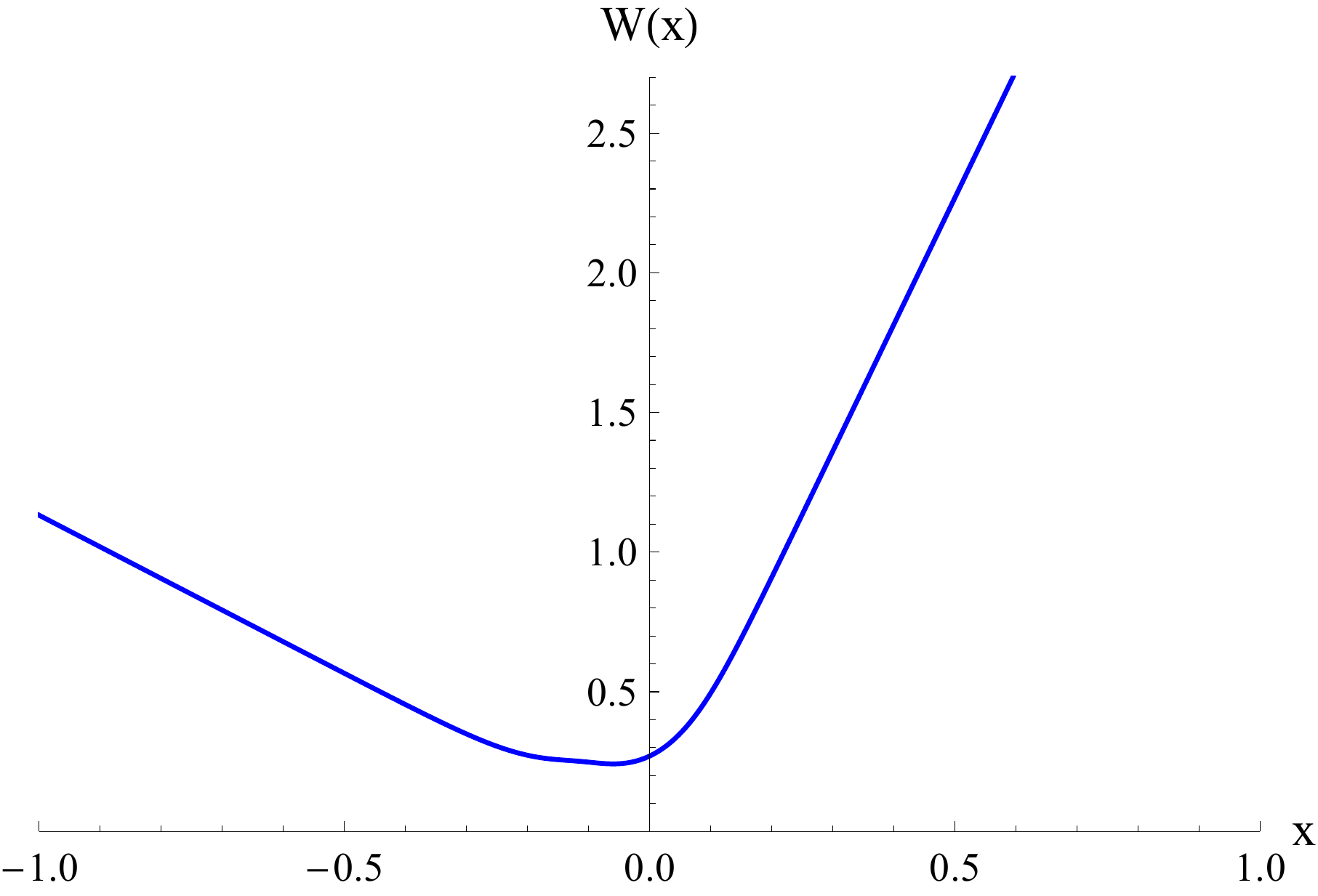}}
\caption{The distribution of the  wormhole configuration as a function of $x$  with the parameters given in the last row of Tab. \ref{tabl1}.  In the top left panel, the purple and red lines denote the functions $F(x)$ and $G(x)$, respectively.  In the bottom left panel,  the red point corresponds to the minimum value of $N(x)$ with $N'(x=-0.0815)=0$. }\label{fig1}
\end{figure}

It is worth pointing out that there may exist various possible solutions for a given set  of
parameters. In fact, these solutions can be considered as the ground or excited {states of the wormhole}, where the Dirac field {possessing zero and $n(>0)$} nodes along the radial coordinate corresponds to the ground state and the $n$-th excited state, respectively.

   \begin{large}
 \textbf{ Acknowledgments}
  \end{large}
We thank R. A. Konoplya and A. Zhidenko
for sharing their numerical data and helpful discussions.
This work was supported by National Key Research and Development Program of China (Grant No. 2020YFC2201503), the National Natural Science Foundation of China (Grants No. 11875151, 12075103, and No.~12047501), the 111 Project under (Grant No. B20063) and ``Lanzhou City's scientific research funding subsidy to Lanzhou University".

\end{document}